\documentclass[aps,prl,reprint,showpacs]{revtex4-1}
\usepackage{amsmath}
\usepackage{SIunits}
\usepackage{graphicx}

\begin{document}

\newcommand{\A}[0]{\ensuremath{\hat{A}}}
\newcommand{\Adag}[0]{\ensuremath{\hat{A}^{\dagger}}}
\newcommand{\bra}[1]{\ensuremath{\left<#1\right|}}
\newcommand{\brae}[0]{\ensuremath{\left<\mathrm{e}\right|}}
\newcommand{\brag}[0]{\ensuremath{\left<\mathrm{g}\right|}}
\newcommand{\brai}[0]{\ensuremath{\left<\mathrm{i}\right|}}
\renewcommand{\c}[0]{\ensuremath{\hat{c}}}
\newcommand{\cdag}[0]{\ensuremath{\hat{c}^{\dagger}}}
\newcommand{\dens}[0]{\ensuremath{\rho}}
\newcommand{\densc}[0]{\ensuremath{\rho}_{\mathrm{c}}}
\newcommand{\denspast}[0]{\ensuremath{\rho_{\mathrm{p}}}}
\newcommand{\densee}[0]{\ensuremath{\hat{\rho}^{\mathrm{ee}}}}
\newcommand{\denseg}[0]{\ensuremath{\hat{\rho}^{\mathrm{eg}}}}
\newcommand{\densge}[0]{\ensuremath{\hat{\rho}^{\mathrm{ge}}}}
\newcommand{\densgg}[0]{\ensuremath{\hat{\rho}^{\mathrm{gg}}}}
\newcommand{\densext}[0]{\ensuremath{\hat{\rho}^{\mathrm{ext}}}}
\newcommand{\densS}[0]{\ensuremath{\hat{\rho}^{\mathrm{S}}}}
\newcommand{\E}[0]{\ensuremath{E}}
\newcommand{\gammap}[0]{\ensuremath{\gamma_{\mathrm{p}}}}
\newcommand{\gammapar}[0]{\ensuremath{\gamma_{\parallel}}}
\newcommand{\gammaperp}[0]{\ensuremath{\gamma_{\perp}}}
\renewcommand{\H}[0]{\ensuremath{\hat{H}}}
\renewcommand{\Im}[0]{\ensuremath{\mathrm{Im}}}
\newcommand{\JS}[0]{\ensuremath{\mathcal{J}_{\mathrm{S}}}}
\newcommand{\ket}[1]{\ensuremath{\left|#1\right>}}
\newcommand{\kete}[0]{\ensuremath{\left|\mathrm{e}\right>}}
\newcommand{\ketg}[0]{\ensuremath{\left|\mathrm{g}\right>}}
\newcommand{\keti}[0]{\ensuremath{\left|\mathrm{i}\right>}}
\renewcommand{\L}[0]{\ensuremath{\hat{L}}}
\newcommand{\Ldag}[0]{\ensuremath{\hat{L}^{\dagger}}}
\newcommand{\mat}[1]{\ensuremath{\mathbf{#1}}}
\newcommand{\mean}[1]{\ensuremath{\langle#1\rangle}}
\newcommand{\pauli}[0]{\ensuremath{\hat{\sigma}}}
\newcommand{\Pdown}[0]{\ensuremath{\hat{P}_{\mathrm{down}}}}
\newcommand{\Pup}[0]{\ensuremath{\hat{P}_{\mathrm{up}}}}
\renewcommand{\Re}[0]{\ensuremath{\mathrm{Re}}}
\newcommand{\Tr}[0]{\ensuremath{\mathrm{Tr}}}
\renewcommand{\vec}[1]{\ensuremath{\mathbf{#1}}}

\newcommand{\id}{\ensuremath{\hat I}}

\title{Past quantum states}
\author{S\o ren Gammelmark}
\author{Brian Julsgaard}
\author{Klaus M{\o}lmer}
\email{moelmer@phys.au.dk}
\affiliation{Department of Physics and Astronomy, Aarhus University, Ny
  Munkegade 120, DK-8000 Aarhus C, Denmark.}

\date{\today}

\begin{abstract}
  A density matrix $\dens(t)$ yields probabilistic information about
  the outcome of measurements on a quantum system. We introduce here
  the past quantum state, which, at time $T$, accounts for the state
  of a quantum system at earlier times $t < T$. The past quantum state
  $\Xi(t)$ is composed of two objects, $\dens(t)$ and $\E(t)$,
  conditioned on the dynamics and the probing of the system until $t$
  and in the time interval $[t,T]$, respectively. The past quantum
  state is characterized by its ability to make better predictions for
  the unknown outcome of projective and weak value measurements at $t$
  than the conventional quantum state at that time. On the one hand,
  our formalism shows how smoothing procedures for estimation of past
  classical signals by a quantum probe [M.~Tsang,
  Phys.~Rev.~Lett.{\bf~102}, 250403, (2009)] apply also to describe
  the past state of the quantum system itself. On the other hand, it
  generalizes theories of pre- and post-selected quantum states
  [Y.~Aharonov and L.~Vaidman, J.~Phys.~A: Math.~Gen.{\bf~24}, 2315
  (1991)] to systems subject to any quantum measurement scenario, any
  coherent evolution, and any Markovian dissipation processes.
\end{abstract}


\maketitle

\begin{figure*}[t]
  \centering
  \includegraphics{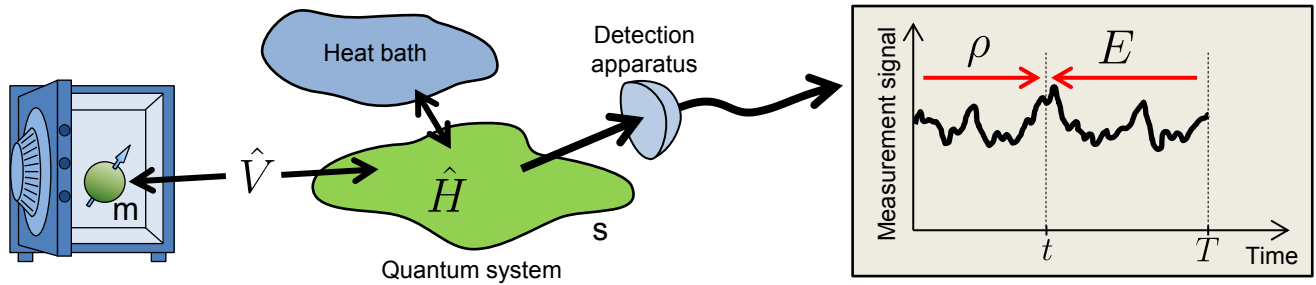}
  \caption{Modeling a past measurement: A quantum system ``s'' is
    monitored during $[0,T]$ and possibly subject to coherent
    evolution by the Hamiltonian $\hat{H}$ and coupling to a heat
    bath. At time $t$ a physical property of s is mapped to a meter
    ``m'' using an interaction $\hat{V}$. This meter, or the outcome
    of a measurement performed on the meter, is stored in a safe
    immediately after the interaction at time $t$.  The matrices
    $\dens$ and $\E$, depending on the measurement signal before and
    after $t$, respectively, constitute the \emph{past quantum state},
    which predicts \emph{better} the measurement outcome than $\dens$
    alone when the safe is opened at time $T$.}
  \label{fig:Setup}
\end{figure*}

Quantum systems are described by wave functions or density matrices,
which yield probabilistic predictions for the outcome of measurements
performed on the systems. Following upon the rules laid out with the
foundations of quantum theory in the 1920'es, the description of
measurements on a so-called open quantum system has in the last few
decades evolved into a well-established stochastic theory
\cite{Wiseman.QuantumMeasurementControl}. According to this theory,
the density matrix $\dens(t)$ evolves with time in a manner governed,
on the one hand, by the system Hamiltonian and damping terms and, on
the other hand, by the back action associated with the random outcome
of measurements performed on the system or its environment. In this
article, we introduce a new element in the quantum description of
probed quantum systems: the past quantum state. While the density
matrix $\dens(t)$ yields predictions about the outcome of the
measurement of any observable at time $t$ conditioned on previous
measurements, the past quantum state yields better predictions for the
same measurement by being conditioned on all measurements carried out
until the present time.  The past quantum state is the state that we,
based on what we know now, assign to a quantum system in the past. It
is thus similar to the completely natural assignment of probabilities
to past values of classical random quantities, e.g., for a Brownian
particle detected at position $x$ to have been at the position $y$ at
given earlier times.  Here, we provide a generalization of the
assignment of probabilities to past classical stochastic processes to
the quantum case. Along with the definition and derivation of a past
quantum state formalism, we shall answer the pressing questions: What
does it mean to make predictions about the past?  What are the new
results and applications of a theory of past quantum states?

Consider an open quantum system subject to our continuous probing as
illustrated in Fig.~\ref{fig:Setup}. We assume the initial quantum
state $\dens_0$, at time $t = 0$, and we probe the system until time
$T$ such that, conditioned on the measurement outcomes, the density
matrix $\dens(t)$ is given at any time $t \in [0,T]$. Specifically, if
a different observer performs a measurement on the system at the
intermediate time $t$, the density matrix $\dens(t)$ provides the
probability distribution of the possible measurement outcomes. Probing
of the system after time $t$ yields results that further refine our
knowledge about the system at $t$ and, indeed, there exists an
\emph{effect matrix}, $\E(t)$, assuming the same Hilbert space
dimension as $\dens(t)$, which depends on the dynamics and on the
information acquired later than $t$ until the present time $T$ such
that the pair of matrices,
\begin{equation}
\label{def:past_quantum_state}
  \Xi(t)=(\dens(t),\,\E(t)),
\end{equation}
together enable better predictions than $\dens(t)$ alone for the
outcome of measurements carried out at time $t$. To discuss in a
meaningful way what is meant by predicting a past measurement, we
consider the setup shown in Fig.~\ref{fig:Setup}. Through an
appropriately chosen interaction the physical property of interest is
extracted at time $t$ via coupling to another quantum system, an
ancillary ``meter''. This meter is stored ``in a safe'', or it may be
immediately measured and the result stored for later inspection. We
show that $\Xi(t)$ provides better predictions than $\rho(t)$ of what
will eventually be observed when the safe is opened. This qualifies
$\Xi(t)$ rather than $\rho(t)$ to be associated to the \emph{past
  quantum state} of the system.

The text book description of projective quantum measurements of a
Hermitian operator is a specific case of general measurements
associated with the action of different operators $\hat{\Omega}_m$
that fulfill $\Sigma_m \hat{\Omega}^{\dagger}_m \hat{\Omega}_m = \id$,
where $\id$ is the identity and $m$ is an index referring to the
possible measurement outcomes \cite{Nielsen.QuantCompInf}. For such a
generalized measurement a suitable generalization of Born's rule
provides the probability at time $t$ for observing the outcome $m$:
$p(m) = \Tr(\hat{\Omega}_m \dens(t) \hat{\Omega}^{\dagger}_m)/
\Tr(\dens(t))$. In the Supplementary Information \cite{SuppMater} we
prove that after further probing of the system until time $T$, the
probability that the outcome is $m$ depends on $\Xi(t)$, i.e., on both
$\rho(t)$ and $\E(t)$:
\begin{equation}
  \label{eq:p(m)_past}
    p_{\mathrm{p}}(m) = \frac{\Tr(\hat{\Omega}_m\dens(t)\hat{\Omega}^{\dagger}_m
     \E(t))}{\sum_m\Tr(\hat{\Omega}_m\dens(t)\hat{\Omega}^{\dagger}_mE(t))}.
\end{equation}
This formula is general and covers all possible measurement scenarios
and any Markovian dynamical evolution, observed or non-observed, of
our quantum system. As exemplified below and derived formally in the
Supplementary Information \cite{SuppMater}, $\E(t)$ can be calculated
backward in time following an \emph{adjoint} equation very similar to
the forward evolution of $\dens(t)$. In absence of probing, $\E$
retains its value $\E = \id$ for all times, and
Eq.~(\ref{eq:p(m)_past}) reduces to the conventional expression since
only our observations can further the knowledge of the state.

In the special case of a projective measurement of an observable
$\hat{A}$, Eq.~(\ref{eq:p(m)_past}) applies with $\hat{\Omega}_m =
\hat{\Pi}_m$ denoting orthogonal projection operators on the
eigenstates of $\hat{A}$. Past mean values, variances, and higher
moments of $\hat{A}$ then follow in the usual manner from
$p_{\mathrm{p}}(m)$.  It is interesting to note that variances of past
measurement outcomes will not necessarily obey Heisenberg's
uncertainty relations for non-commuting operators $\hat{A}$ and
$\hat{B}$. This, however, is not in violation of quantum mechanics
since our probabilistic statements concern only the value of the
observable actually measured, i.e., either $\hat{A}$ or $\hat{B}$. In
Ref.~\cite{Vaidman.PhysRevLett.58.1385(1987)}, a clever guessing game
is suggested, where a person is given a spin-1/2 particle, and is free
to measure either the $x$, $y$ or $z$ component of the spin, and
subsequently return the particle. By preparing the initial state and
measuring the final state, it is possible to announce the outcome of
any of these measurements without uncertainty.  The past quantum state
also permits a full analysis of such games.

Our general formalism permits also an analysis of so-called weak value
measurements \cite{Aharonov.PhysRevLett.60.1351(1988)}. In this case
the strength of the interaction between a system observable $\hat{A}$
and the meter can be parametrized by a small number $\epsilon \ll 1$,
such that the disturbance resulting from the measurement is
proportional to $\epsilon^2$ and thus may be neglected. Nonetheless,
averaged over sufficiently many experimental realizations the meter
read-out will reveal the \emph{mean value} of the system observable
$\hat{A}$ by the formula: $\mean{\hat{A}}_{\mathrm{w}} =
\Tr(\hat{A}\denspast)$ where we have defined the \emph{past density
  matrix} $\denspast = \dens\E/\Tr(\dens\E)$ \cite{SuppMater}. Here
$\dens$ and $\E$ are exactly the constituents of the past quantum
state~(\ref{def:past_quantum_state}) and we thus provide a
generalization of existing weak value expressions:
$\mean{\hat{A}}_{\mathrm{w}} = \Re\{\bra{\varphi} \A \ket{\psi}
/\bra{\varphi}\psi\rangle\}$ or $\mean{\hat{A}}_{\mathrm{w}} =
\Re\{\Tr(\A\dens_i \E_m)/\Tr(\dens_i\E_m)\}$, which apply,
respectively, for a system initially prepared in a pure state
$\ket{\psi}$ or mixed state $\dens_i$ and subsequently projected into
the final state $\ket{\varphi}$ or detected by a generalized
measurement operator $\E_m$ \cite{Aharonov.PhysRevLett.60.1351(1988),
  Johansen.PhysRevA.70.052115(2004)}. In these two examples
$\denspast$ takes the form
$\ket{\psi}\bra{\varphi}/\bra{\varphi}\psi\rangle$ or $\dens_i
\E_m/\Tr(\dens_i\E_m)$, which has recently led to the recognition of
these expressions as pre- and post-selected ``connection states''
\cite{Kofman.arXiv.1303.6031(2013)}. In our general theory, the
quantum state may at any time in the past be viewed as a pre-selection
$\dens$ by earlier measurement and a post-selection $\E$ by the later
ones, with unitary evolution, dissipation processes, and direct
measurement on the system also accounted for.  We note that
$\denspast$ is not described by a single evolution equation, but
$\dens$ and $\E$ must be calculated separately.

\begin{figure}[t]
  \centering
  \includegraphics{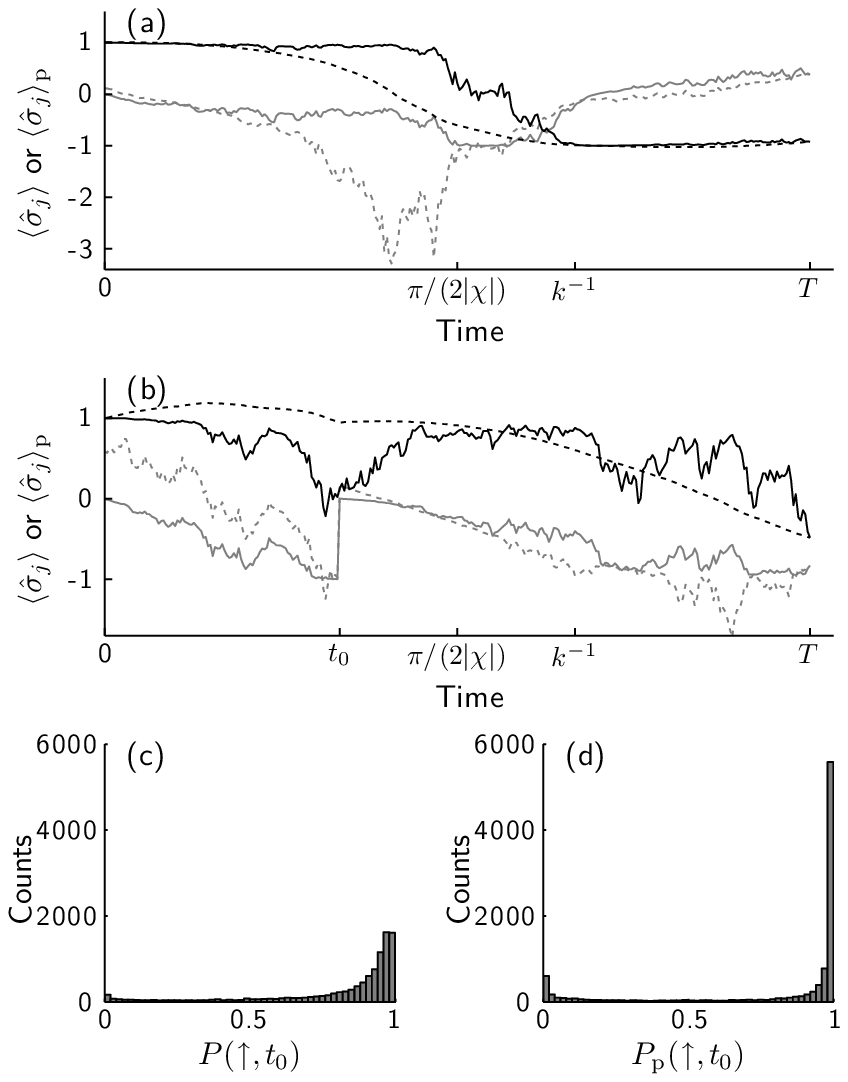}
  \caption{Forward and past expectation values of a continuously
    monitored spin-1/2 particle: In panels (a) and (b) the curves
    show: $\mean{\pauli_z}$ [solid black], $\mean{\pauli_x}$ [solid
    gray], $\Re\{\mean{\pauli_z}_{\mathrm{p}}\}$ [dashed black], and
    $\Re\{\mean{\pauli_x}_{\mathrm{p}}\}$ [dashed gray]. In panel (b)
    an observer has performed a projective measurement of $\pauli_z$
    on the system at time $t_0$ without revealing the result. This
    leads to $\mean{\pauli_x}$ resetting to zero and the spectral
    boundary $-1 \le \Re\{\mean{\pauli_z}_{\mathrm{p}}\} \le 1$
    automatically enforced. Panels (c) and (d) show the probability
    distribution for this projective measurement revealing spin-up
    based on $\dens(t_0)$ and $\Xi(t_0)$, respectively, using 10,000
    simulations.}
  \label{fig:Graphs}
\end{figure}

To illustrate the past state formalism and some of its results for a
physical problem, let us turn to an example with a quantum system
subject to coherent evolution, dissipation, and continuous
homodyne-like monitoring. In such problems the usual quantum state
$\dens(t)$ is conditioned on the detection record until $t \in [0,T]$
and it formally obeys the corresponding stochastic master equation
\cite{Wiseman.QuantumMeasurementControl,
  Jacobs.ContemporaryPhys.47.279(2006)}:
\begin{equation}
\label{eq:Forward}
\begin{split}
   d\dens_t = &-i[\H,\dens_t]dt + \sqrt{\eta}(\c\dens_t + \dens_t\cdag)dY_t\\
   & + \sum_m (\L_m\dens_t\Ldag_m - \frac{1}{2}\{\Ldag_m\L_m,\dens_t\})dt,
\end{split}
\end{equation}
where $d\dens_t = \dens_{t+dt} - \dens_t$, $\H$ is the interaction
Hamiltonian, $\c$ is the measurement observable, $\L_m$ are Lindblad
operators describing the dissipative coupling to the environment (including $\c$
as one of the $\L_m$-terms), $\eta$ is the detector efficiency, and
$dY_t$ is the measurement record. The effect matrix $\E$ solves a
corresponding adjoint equation with final condition $\E = \id$ at $T$
\cite{SuppMater}:
\begin{equation}
\label{eq:Backward}
\begin{split}
  d\E_t = &i[\H,\E_t]dt + \sqrt{\eta}(\cdag\E_t + \E_t\c)dY_{t-dt}\\
      & + \sum_m(\Ldag_m\E_t\L_m - \frac{1}{2}\{\Ldag_m\L_m,\E_t\})dt,
\end{split}
\end{equation}
where $dt$ is positive and $d\E_t \equiv \E_{t-dt} - \E_t$,
propagating backward from $T$ to $t$ using the same measurement record
$dY_t$ as in Eq.~(\ref{eq:Forward}). We note that these equations are
not trace preserving but can easily be adapted as such if required,
e.g.~for numerical evaluation.

For concreteness, we consider now a quantum two-level system subject
to coherent driving with Rabi frequency $\chi$ according to the
Hamiltonian $\H = \frac{1}{2}(\chi\pauli_+ + \chi^*\pauli_-)$ and to
continuous probing of $\pauli_z$ through the measurement operator $\c
= \sqrt{k}\pauli_z$. Here $\pauli_j$ are Pauli spin operators and $k$
is the measurement strength. Such measurement could be implemented by,
e.g., polarization rotation of a radiation field coupled to the
spin-1/2 particle \cite{Takahashi.PhysRevA.60.4974(1999)}. We have
performed simulations with $\eta = 1$, a pure initial state $\dens_0 =
\ket{\uparrow}\bra{\uparrow}$, and an imaginary $\chi$ such that the
coherent driving rotates the spin around the $y$-axis. In
Fig.~\ref{fig:Graphs}, the expectation value of the Pauli operators,
$\pauli_z$ and $\pauli_x$, for the quantum system are compared using
the usual forward state $\dens(t)$ and the past quantum state
$\Xi(t)$.

The analysis in Fig.~\ref{fig:Graphs}(a) applies to the case where the
system has not been disturbed by further measurements by other
observers. This implies that the past density matrix
$\dens_{\mathrm{p}}(t)$ may be used to yield predictions for the
outcome of weak value measurements of any system observable.  We
stress that despite its possible values beyond the interval $[-1,1]$,
$\mean{\pauli_j(t)}_{\mathrm{p}}= \Tr(\pauli_j\denspast(t))$, rather
than $\mean{\pauli_j(t)} = \Tr(\pauli_j\dens(t))$, represents the
correct estimate of the disturbance of the meter system. For a
spin-$1/2$ meter system, the real and imaginary parts of
$\mean{\pauli_j(t)}_{\mathrm{p}}$ correspond to mean rotation angles
of the spin around different axes
\cite{Shengjun.PhysLettA.374.34(2009)}.

Fig.~\ref{fig:Graphs}(b) exemplifies the case where an observer has
performed a projective measurement of $\pauli_z$ at time $t_0$ without
revealing the result, and Eq.~(\ref{eq:p(m)_past}) enables a past
prediction $\mean{\pauli_z(t_0)}_{\mathrm{p}}$ of this outcome using
$\Xi(t_0)$. For all other times $t \in [0,T]$ the projective
measurement must be taken into account in the evolution of $\dens$ and
$\E$. To account for the decoherence by the measurement at $t_0$, we
evolve the density matrix, $\dens(t_{0+}) = \Pup\dens(t_{0-})\Pup +
\Pdown\dens(t_{0-})\Pdown$ by the projection operators, $\Pup =
\ket{\uparrow}\bra{\uparrow}$ and $\Pdown =
\ket{\downarrow}\bra{\downarrow}$. Similarly, to obtain the value of
the effect matrix $\E$ prior to $t_0$, we have to apply the operation
$\E(t_{0-}) = \Pup\E(t_{0+})\Pup + \Pdown\E(t_{0+})\Pdown$. It is
particularly interesting to compare the predictions of the un-revealed
measurement outcome using the conventional and the past quantum state
formalism. For predictions associated with projective measurements,
$\mean{\pauli_z(t_0)}_{\mathrm{p}}$ is real and it remains within its
spectral boundaries as it should to yield agreement with
experiments. The result, however, differs from the prediction by the
conventional density matrix, and we quantify this difference by the
distribution of probabilities for the two-level spin direction to be
registered as up or down. These distributions are shown for the
conventional quantum state in Fig.~\ref{fig:Graphs}(c) and for the
past quantum state in Fig.~\ref{fig:Graphs}(d). By assuming that the
most likely measurement result is the one occurring, we guess the
outcome correctly with 88\% probability by the conventional quantum
state $\rho(t_0)$, while the past quantum state $\Xi(t_0)$ yields the
correct measurement outcome with $94\%$ probability.

We note that if $\dens(t)$ predicted a measurement outcome with
certainty at time $t$, our later probing will never lead to
disagreement with this prediction: If the system was in the $m$th
eigen state $\ket{a_m}$ of an observable $\A$, i.e.~$\A\ket{a_m} =
a_m\ket{a_m}$, at time $t$, then $\dens(t) = \ket{a_m}\bra{a_m} =
\hat{\Pi}_m$ and the past probability for measuring the $n$th eigen
state then becomes: $p_{\mathrm{p}}(n) = \delta_{n,m}$ according to
Eq.~(\ref{eq:p(m)_past}) and the orthogonality relation
$\hat{\Pi}_m\hat{\Pi}_n = \hat{\Pi}_m\delta_{n,m}$.

While the past quantum state enables a sharper prediction for the
projective measurement shown in Fig.~\ref{fig:Graphs}(c,d), we note
that the weak value $\mean{\pauli_z}_{\mathrm{p}}$ is actually
smoother than the forward estimate $\mean{\pauli_z}$ in
Fig.~\ref{fig:Graphs}(a,b). In the Supplementary Information we show
that while $\mean{\pauli_z}$ varies with noisy increments $\propto
dY_t \propto \sqrt{dt}$ due to Eq.~(\ref{eq:Forward}), the changes in
$\mean{\pauli_z}_{\mathrm{p}}$ at time $t$ are formally independent of
$dY_t$ and vary smoothly according to the integrated noise via
$\rho(t)$ and $E(t)$.

\begin{figure}[t]
  \centering
   \includegraphics{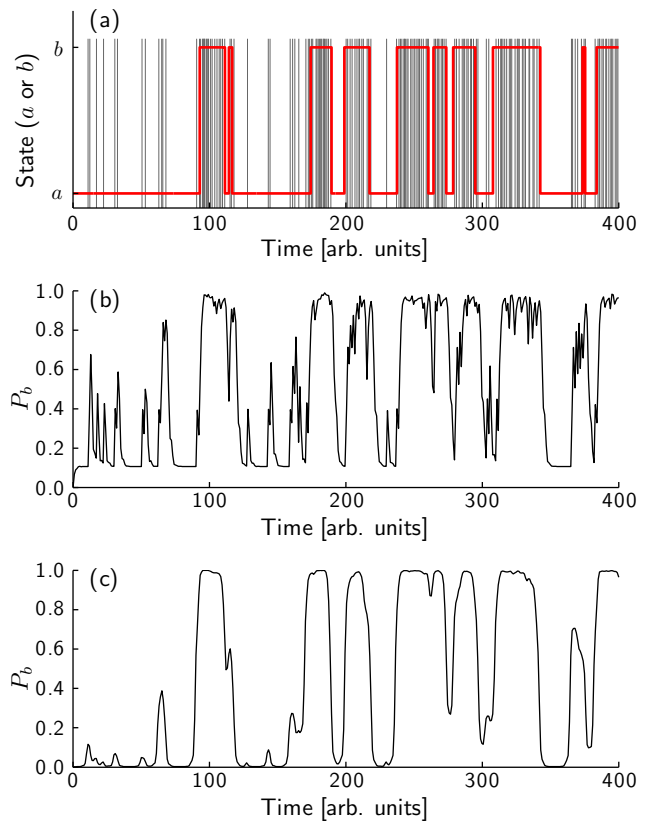}
  \caption{Forward and past estimate of discrete position jumps of a
    driven two-level atom: In panel (a), the actual random position
    jumps of the atom are shown (red line) along with the photon
    detection times (vertical black).  At position $a$ ($b$), the
    atoms have a low (high) average emission rate.  Panel (b) shows
    the probability for finding the atom at site $b$ calculated using
    the forward density matrix, while panel (c) employs the past
    quantum state using the same data. Panel (c) is clearly more
    decisive and in much better agreement with the true state (red
    line in panel (a)).}
  \label{fig:SmoothingExample}
\end{figure}

To exemplify an estimation process with mixed coherent and incoherent
degrees of freedom, we consider in Fig.~\ref{fig:SmoothingExample} a
coherently driven, spontaneously decaying two-level atom which can
jump incoherently between two sites, $a$ and $b$. In
Fig.~\ref{fig:SmoothingExample}(a) the position of the atom (as used
in the simulations but in experimental realizations hidden from us) is
shown along with the instants of detection of photons emitted from the
two-level atom. Owing to different site environments, the coherent
atom-dynamics is determined by different parameter values at each
site.  We can use the forward and past quantum states, based on the
photo-detection record only, to estimate the location of the atom as
shown in Fig.~\ref{fig:SmoothingExample}(b) and
Fig.~\ref{fig:SmoothingExample}(c), respectively. As demonstrated in
Refs.~\cite{Tsang.PhysRevLett.102.250403(2009),
  Tsang.PhysRevA.80.033840(2009), Tsang.PhysRevA.81.013824(2010),
  Tsang.PhysRevLett.106.090401(2011)}, a similar formalism enables the
estimation of unknown classical perturbations, applied to a probed
quantum system. These works generalize the so-called smoothing
procedure applied in the classical probability theory of Hidden Markov
Models (HMM) \cite{NumericalRecipesThirdEd} to hybrid quantum
classical systems. Such hybrid systems may, indeed, be embedded in a
full quantum model, and we comment further on the formal similarities
between our past quantum state and the smoothing procedure in
classical HMM in the Supplementary \cite{SuppMater}.  The past quantum
state estimate is both much less noisy and much more decisive,
enabling a more accurate state estimation and, e.g., a better estimate
of the $a$-$b$ jump rate. In addition to estimating such classical
properties our formalism provides also a better estimation of the
quantum system itself and we are currently investigating the
application of our theory to the state of a photon field in a cavity
\cite{Gleyzes.Nature.446.297(2007)} and to the state of a
superconducting qubit \cite{Hatridge.Science.339.178(2013),
  Vijay.PhysRevLett.106.110502(2011), Groen.arXiv.1302.5147(2013)}.

The past quantum state $\Xi(t)$ depends on events occurring in the
future beyond the time $t$. While ``spooky action from the future''
via post-selection has stimulated fascinating scientific debate
\cite{Aharonov.PhysToday.63.27(2010)}, the predictions we make can be
interpreted as correlations between system observables at the past
time $t$ and a probing signal acquired in the, also past, interval
$[0,T]$. This distinction has also been made clear in the formal work
on pre- and post-selected states
\cite{Aharonov.JPhysAMathGen.24.2315(1991),
  Aharonov.PhysRevA.79.052110(2009)}. While the concept of a past
quantum state is also central in these publications, our analysis aims
less on the foundations of quantum theory and more on the general and
explicit description of continuously monitored open quantum systems that are
commonplace in laboratories today. Our theory applies to experiments
on superconducting devices
\cite{Palacios-Laloy.NaturePhys.6.442(2010),
  Vijay.PhysRevLett.106.110502(2011), Hatridge.Science.339.178(2013),
  Groen.arXiv.1302.5147(2013)}; semiconductor quantum dots
\cite{Vink.NaturePhys.5.764(2009)}; NV-centers in diamond
\cite{Waldherr.PhysRevLett.107.090401(2011),
  Cai.NewJPhys.15.013020(2013)}; nuclear spins in silicon
\cite{Pla.Nature.496.334(2013)}; trapped ions, atoms and molecules
\cite{Kirchmair.Nature.460.494(2009), Kubanek.Nature.462.898(2009),
  Basche.Nature.373.133(1995)}; photons
\cite{Gleyzes.Nature.446.297(2007), Goggin.PNAS.108.1256(2011),
  Kocsis.Science.332.1170(2011)}, and nanomechanical devices
\cite{Naik.Nature.443.193(2006)}. These systems, indeed, may be used
for fundamental tests \cite{Palacios-Laloy.NaturePhys.6.442(2010),
  Waldherr.PhysRevLett.107.090401(2011),
  Kirchmair.Nature.460.494(2009), Goggin.PNAS.108.1256(2011),
  Groen.arXiv.1302.5147(2013)}, but they also hold the potential for
application in precision probing
\cite{Giovannetti.Science.306.1330(2004)} and quantum information
science \cite{Nielsen.QuantCompInf}, and we hope our theory will
stimulate further development of both technical and foundational
aspects of the theory of open quantum systems.

This work was supported by the Villum Foundation.


%

\end{document}


\title{Supplementary Information}
\author{S\o ren Gammelmark}
\author{Brian Julsgaard}
\author{Klaus M{\o}lmer}

\affiliation{Department of Physics and Astronomy, Aarhus University, Ny
  Munkegade 120, DK-8000 Aarhus C, Denmark.}
\date{\today}

\maketitle



In this supplementary note we show that with our definition of the past quantum state we obtain identical results for past measurement outcomes as predicted by the ordinary quantum formalism, when this is applied to deferred measurements on a combined system-meter set-up \cite{Griffiths.PhysRevLett.76.3228(1996)}.
Instead of an expression for determining past measurement outcomes propagating forward in time and including the meter system, we derive a backward propagating \emph{effect matrix} $E$ for the system only.
We provide the specific stochastic differential equations for diffusion and jump type probing of the system (corresponding to homodyne detection and photon counting schemes). These equations can be chosen linear and non-trace preserving or non-linear and trace-preserving as the usual quantum filtering equations \cite{Wiseman.QuantumMeasurementControl}.
Finally, we demonstrate that our past quantum state generalizes the so-called \emph{smoothed} state in classical hidden Markov models to the quantum case:
The conditional density matrix $\rho(t)$ is a natural generalization of the $\alpha$-state conditional probability, and the backward propagating effect matrix
$E$ is the quantum generalization of the backward propagating $\beta$-state in hidden Markov models \cite{NumericalRecipesThirdEd}.

\chapter{Deferred measurements and the past quantum state}

\begin{figure}
 \centering
 \includegraphics[width=\columnwidth]{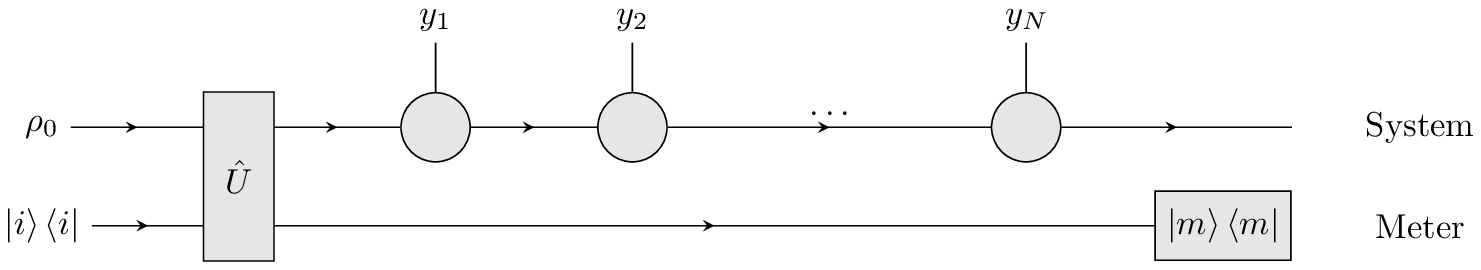}
 \caption{Graphical illustration of the system dynamics considered in this note.
Our open quantum system (top line) and a meter (bottom line) interact as described by the unitary operator $\hat U$.
The system is probed by subsequent measurements with outcomes $y_1, \ldots y_N$, and later, the meter is
subject to a projective measurement in some basis $\{\ket m\}$.
 }
  \label{fig:MeterSystemDiagram}
\end{figure}

Imagine an observer using a meter to perform any measurement on our system which is initially prepared in a state which is represented by the density matrix $\rho_0$.
The observer correlates the meter and our system by a unitary interaction, $\hat U$, between them.
After this unitary interaction the system evolves independently of the meter and during this evolution, we measure our system a number of times $N$ as illustrated in Fig. \ref{fig:MeterSystemDiagram}.
By taking a suitable limit with $N\to\infty$ a description of continuous-time observation can be obtained.
The observer who now has the meter in her possession can choose to perform a measurement of her choice at any later time.
Our goal is to predict the result of such a measurement.

We have access to the results of the $N$ measurements, where each measurement is in the most general case described by a set of measurement effect operators $\hSystem_y$ for each measurement outcome $y$ in the set of possible measurement results $Y$.
Using this formalism the effect of the measurement result $y$ is to update a density matrix $\rho$ according to $\rho \overset{y}{\mapsto} \hSystem_y \rho \hSystem_y^\dagger / \trace(\hSystem_y^\dagger \hSystem_y \rho)$.
The resulting density matrix, which we denote $\rho|y$, is conditioned upon the measurement result $y$.
The probability of obtaining the result $y$ is given by $P(y) = \trace(\hSystem_y^\dagger \hSystem_y \rho)$.

The operators $\hSystem_y$ should satisfy $\sum_{y\in Y} \hSystem_y^\dagger \hSystem_y = \id$ such that the probability for observing any $y\in Y$ is unity.
If the observation is made with less than 100\% readout efficiency the resulting conditioned density matrix can be written as a sum
\begin{align}
 \rho \overset{y}{\longmapsto} \frac{ \sum_{k=1}^K \hSystem_{k|y} \rho \hSystem_{k|y}^\dagger } { \trace\left( \sum_{k=1}^K \hSystem_{k|y}^\dagger \hSystem_{k|y} \rho \right) }, \label{eq:QuantumOperation}
\end{align}
where $\hSystem_{k|y}$ are operators describing different possible effects which are all associated with the measurement result $y$. The number of terms, $K$, can depend on $y$.

Equation (\ref{eq:QuantumOperation}) captures the effect of most types of dynamics of open quantum systems, including unitary evolution, dissipation effects described by a master equation on Lindblad form, and measurements.
A density matrix subject to unitary evolution with unitary operator $\hat U$ is thus updated according to $\rho \mapsto \hSystem \rho \hSystem^\dagger$. Where only the single unitary operator $\hSystem$ is needed clearly satisfies $\hSystem^\dagger \hSystem = \id$.

A projective measurement is described by a set of projectors $\hat \Pi_a$ where $a$ are eigenvalues of the observable $\hat A$ being measured, and this case is also included by the identification $\hSystem_a = \hat \Pi_a$.
The effect on the system density matrix by a projective measurement is then simply the usual projection postulate $\rho \overset{a}{\longmapsto} \hat \Pi_a \rho \hat \Pi_a / \trace(\hat \Pi_a \rho)$.
If we know that a projection measurement is performed, but the result of the measurement is hidden from us, the density matrix is updated according to $\rho \longmapsto \sum_a \hat \Pi_a \rho \hat \Pi_a$.
In this final example, all off-diagonal density matrix elements in the eigen basis of $\hat A$ are zeroed, and the unobserved measurement operation is therefore equivalent to a decoherence process.

If a number of measurements are performed in sequence the density matrix $\rho$ is updated repeatedly by the formula \refeq{eq:QuantumOperation}.
In the limit of continuous-time measurements we can also describe the conditional time-evolution of the system density matrix.
Consider for example a quantum system subject to homodyne detection.
In this case the effect of a detection in a small interval of time $dt$ is given by the operators
\begin{align}
 \hSystem_{dY_t} = (2\pi dt)^{-1/4} \exp(-dY_t^2/4 dt) (\id - i \hat H dt - \hat c^\dagger \hat c / 2 dt + \hat c dY_t), \label{eq:DiffusionMeasurementOperator}
\end{align}
where $\hat H$ is the system Hamiltonian and $\hat c$ is the system operator coupling to the real homodyne output signal $dY_t$. \cite{Wiseman.QuantumMeasurementControl,Jacobs.ContemporaryPhys.47.279(2006)}
The probability for observing $dY_t$ in this infinitesimal interval of time is a normal distribution with mean value $\trace((\hat c + \hat c^\dagger)\rho) dt$ and variance $dt$.
By applying the update formula \refeq{eq:QuantumOperation} with these operators we get Eq.~(3) with $\eta = 1$ and only one Lindblad operator $\hat L_1 = \hat c$.
By including unobserved output channels for Lindblad operator $\hat L_i$, $i > 1$ and including limited detector efficiency $\eta < 1$, Eq.~(3) turns out to be a special case of \refeq{eq:QuantumOperation}.

Let us return to our main line of inquiry.
The meter is assumed to be initialized in a pure state $\ket i \bra i$ when the observer applies the unitary interaction, acting on the combined system-meter state $\rho_0 \otimes \ket i \bra i$.
The resulting state is $\hat U (\rho_0 \otimes \ket i \bra i) \hat U^\dagger$.
By inserting complete bases of the meter $\id_M = \sum_{m \in M} \ket m \bra m$ we obtain
\begin{align}
 \rho = \sum_{m,m'\in M} \hMeter_m \rho_0 \hMeter_{m'}^\dagger \otimes \ket{m}\bra{m'},
\end{align}
where we have defined the system operators $\hMeter_m = (\id\otimes\bra{m}) \hat U (\id\otimes\ket i)$.
The operators $\hMeter_m$, defined this way, satisfy the requirement for measurement effect operators, $\sum_{m\in M} \hMeter_m^\dagger \hMeter_m = \id$, and with the chosen form for the operators $\hMeter_m$, we use the coupling to the meter to formally interrogate the properties of the open quantum system at the intermediate time $t$.
We imagine that the open system dynamics proceeds, including the observations on the system continue with measurements result $y \in Y$, which cause the continued system dynamics described by \refeq{eq:QuantumOperation}.
In the general formula \refeq{eq:QuantumOperation} we can include unitary dynamics, dissipation channels and partially efficient measurements.
As noted above, all these effects can be included by a suitable choice of the operators $\hSystem_{k|y}$.

Since the subsequent dynamics only concern the system, the operators $\hSystem_{k|y}$ only act on the system degrees of freedom.
The normalized system-meter state conditioned on one measurement with the result $y$ is
\begin{align}
 \rho|y = \frac{ \sum_{k=1}^K \sum_{m,m'\in M} \hSystem_{k|y} \hMeter_m \rho_0 \hMeter_{m'}^\dagger \hSystem_{k|y}^\dagger \otimes \ket{m}\bra{m'} }
 { \sum_{k=1}^K\sum_{m\in M} \trace\left( \hSystem_{k|y} \hMeter_m \rho_0 \hMeter_m^\dagger \hSystem_{k|y}^\dagger \right) }.
\end{align}

We can calculate the expectation value of any meter-observable $\hat X$ in the state $\rho|y$ by the usual formalism,
\begin{align}
 \Expectation[\hat X|y] = \trace(\id\otimes \hat X \rho|y) =
 \frac{ \sum_{k=1}^K \sum_{m,m'\in M} \trace(\hSystem_{k|y} \hMeter_m \rho_0 \hMeter_{m'}^\dagger \hSystem_{k|y}^\dagger) \braket{m'|\hat X|m} }
 { \sum_{k=1}^K\sum_{m'\in M} \trace\left( \hSystem_{k|y} \hMeter_{m'} \rho_0 \hMeter_{m'}^\dagger \hSystem_{k|y}^\dagger \right) }. \label{eq:ConditionalMeterExpectation}
\end{align}

A projective measurement on the meter, which is now conditional on the system measurement result $y$, and yields the result $m$ with the probability
\begin{align}
 P(m|y)
 = \frac{ \sum_{k=1}^K \trace(\hSystem_{k|y} \hMeter_m \rho_0 \hMeter_m^\dagger \hSystem_{k|y}^\dagger) }
 { \sum_{k=1}^K\sum_{m'\in M} \trace\left( \hSystem_{k|y} \hMeter_{m'} \rho_0 \hMeter_{m'}^\dagger \hSystem_{k|y}^\dagger \right) }, \label{eq:ConditionalMeterProbability}
\end{align}
where we have calculated the conditional expectation value of the meter projection operator $\hat X = \ket m \bra m$ to obtain the conditional probability $P(m|y)$.
Note that we can rewrite the numerator in (\ref{eq:ConditionalMeterProbability}) as
\begin{align}
 \sum_{k=1}^K \trace(\hSystem_{k|y} \hMeter_m \rho_0 \hMeter_m^\dagger \hSystem_{k|y}^\dagger) = \trace\left(\hMeter_m \rho_0 \hMeter_m^\dagger \left[ \sum_{k=1}^K \hSystem_{k|y}^\dagger \hSystem_{k|y} \right] \right) \equiv \trace(M_m \rho_0 M_m^\dagger E),
\end{align}
where we have defined the \emph{effect matrix} $E = \sum_{k=1}^K \hSystem_{k|y}^\dagger \hSystem_{k|y}$, and we obtain
\begin{align}
 P(m|y) &= \frac{\trace(\hMeter_m \rho_0 \hMeter_m^\dagger E)}{\trace(\sum_{m'\in M} \hMeter_{m'} \rho_0 \hMeter_{m'}^\dagger E)}. \label{eq:ConditionalMeterProbabilityState}.
\end{align}
This is the probability for the outcomes, stored "in the safe" as described in the main text. \emph{I.e.}, $P(m|y)$ is the retrodicted probability for the result $m$ of a general measurement on the system with the corresponding measurement operator $\{\hMeter_m\}$.
It notably differs from the usual formula $P(m) = \trace(\hMeter_m^\dagger \hMeter_m \rho_0)$, as it is possible to predict the outcome of measurements on the meter better than before we had access to the later measurement result $y$.

In the same way as the usual time dependent quantum state of a system, represented by a wave function $\psi(t)$ or a density matrix $\rho(t)$, yields probabilities for general measurements on the system, we have now identified a mathematical structure, composed of $\rho_0$ and $E$, which provide the probabilities for past measurements on a quantum system. We thus call the pair of matrices $\Xi=(\rho_0, E)$ \emph{the past quantum state}.

If the system has evolved until time $t$, and multiple measurements have been performed before the meter interacts with our system, $\Xi(t)=(\rho(t), E(t))$, where $\rho(t)$ is the usual open system density matrix found by a stochastic equation of evolution conditioned on the measurements \emph{before} time $t$.
If multiple measurements are performed in sequence after the meter has interacted with our system at time $t$, the effect matrix $E(t)$ depends on all measurement results \emph{after} the coupling to the meter.
In the following sections, we will derive efficient equations of evolution to determine $E$ for both general probing scenarios and for a few special cases.


\section{Dynamical equations for the effect matrix}

Assume that the measurements up to time $t$ has been taken into account in the forward state $\rho_0$ then the generalization of \refeq{eq:ConditionalMeterExpectation} when two subsequent measurements with result $y_1$ and $y_2$ are performed is
\begin{align}
 \Expectation[\hat X|y_1, y_2] = \frac{ \sum_{m,m'\in M}\sum_{k_1, k_2=1}^K \trace\left( \hSystem_{k_2|y_2} \hSystem_{k_1|y_1} \hMeter_m \rho_0 \hMeter_{m'}^\dagger \hSystem_{k_1|y_1}^\dagger \hSystem_{k_2|y_2}^\dagger \right) \braket{m'|\hat X|m} }
 {\sum_{m,\in M}\sum_{k_1, k_2=1}^K \trace\left( \hSystem_{k_2|y_2} \hSystem_{k_1|y_1} \hMeter_m \rho_0 \hMeter_{m}^\dagger \hSystem_{k_1|y_1}^\dagger \hSystem_{k_2|y_2}^\dagger \right)}.
\end{align}
By using the cyclic property of the trace, the numerator can be written as
\begin{align}
 \sum_{m,m'\in M} \trace\left(\hMeter_m \rho_0 \hMeter_{m'}^\dagger \sum_{k_1=1}^K \left\{ \hSystem_{k_1|y_1}^\dagger \left[ \sum_{k_2=1}^K \hSystem_{k_2|y_2}^\dagger \hSystem_{k_2|y_2} \right] \hSystem_{k_1|y_1} \right\} \right) \braket{m'|\hat X|m},
\end{align}
In this case, the effect matrix $E$ is therefore given by
\begin{align}
 E = \sum_{k_1=1}^K \hSystem_{k_1|y_1}^\dagger \left[ \sum_{k_2=1}^K \hSystem_{k_2|y_2}^\dagger \hSystem_{k_2|y_2} \right] \hSystem_{k_1|y_1},
\end{align}
where $E$ now depends explicitly on the two future measurement results $y_1$ and $y_2$.

From this we see that the update formula for $E$, as a counterpart to \refeq{eq:QuantumOperation}, is given by the adjoint update
\begin{align}
 E \overset{y}{\longmapsto} \sum_{k=1}^K \hSystem_{k|y}^\dagger E \hSystem_{k|y} \label{eq:AdjointOperation}
\end{align}
where $E$ equals the identity $\id$ at the final time of measurements $T$, and is propagated recursively \emph{backward} as indicated in the case of two measurements,
\begin{align}
 \id \overset{y_2}{\longmapsto} \sum_{k_2=1}^K \hSystem_{k_2|y_2}^\dagger \hSystem_{k_2|y_2} \overset{y_1}{\longmapsto} \sum_{k_1=1}^K \hSystem_{k_1|y_1}^\dagger \left[ \sum_{k_2=1}^K \hSystem_{k_2|y_2}^\dagger \hSystem_{k_2|y_2} \right] \hSystem_{k_1|y_1}.
\end{align}
The propagation is readily generalized to the case of $N$ measurements,
\begin{align}
 E_N = \id \overset{y_N}{\longmapsto} E_{N-1} \overset{y_{N-1}}{\longmapsto} \ldots \overset{y_1}{\longmapsto} E_0.
\end{align}

A hermitian operator $E$ remains hermitian since the right hand side of \refeq{eq:AdjointOperation} is invariant under hermitian conjugation.
Indeed, $E$ has a separate physical interpretation as the positive semi-definite operator which given the state $\rho_0$ yields the probability for the sequence of future measurement results, $P(y_1, \ldots y_N|\rho_0) = \trace(E \rho_0)$.

\section{The past density matrix}

Our definition of the past quantum state necessitates the use of two matrices from which probabilities can be generally determined via
(\ref{eq:ConditionalMeterProbabilityState}). That expression may, however, be simplified in the special, but interesting, case where the system and meter are coupled very weakly.

We choose a two-dimensional quantum system as our meter, represented as a spin-1/2-particle, interacting briefly with our system by an interaction $\hat V = i g (\hat A \hat \sigma^\dagger - \hat A^\dagger \hat \sigma)$ where $\hat \sigma$ is the spin lowering operator and $\hat A$ is \emph{any}, not necessarily hermitian, system operator.
The meter is initially prepared in the spin-down state $\ket{\downarrow}$, and we allow the interaction to be active for a duration $\tau$ such that $\hat U = \exp(\epsilon ( \hat A \hat \sigma^\dagger - \hat A^\dagger \hat \sigma))$ where $\epsilon = \tau g$.
In the (weak) limit $\epsilon \ll 1$
\begin{align}
 \hat U = \id + \epsilon (\hat A \hat \sigma^\dagger  - \hat A^\dagger \hat \sigma ) - \frac{\epsilon^2}{2} \left( \hat A \hat A^\dagger \hat \sigma^\dagger \hat \sigma  + \hat A^\dagger \hat A \hat \sigma \hat \sigma^\dagger \right) + O(\epsilon^3).
\end{align}
If the meter is initialized in spin down in the $z$-direction $\ket{\downarrow}$ then the measurement effect operators in the $z$-basis are given by $\hMeter_\mu = (\id\otimes\bra{\mu}) \hat U (\id\otimes\ket{\downarrow})$ , $\mu=\downarrow,\uparrow$,
\begin{align}
 \hMeter_\downarrow &= \id - \frac{\epsilon^2}{2} \hat A^\dagger \hat A + O(\epsilon^3) \\
 \hMeter_\uparrow &= \epsilon \hat A + O(\epsilon^3).
\end{align}
The effect of the measurement on the system state $\rho_0$ when the result is not revealed is
\begin{align}
 \rho_0 \longmapsto \hMeter_\downarrow \rho_0 \hMeter_\downarrow^\dagger + \hMeter_\uparrow \rho_0 \hMeter_\uparrow^\dagger
  = \rho_0 + O(\epsilon^2).
\end{align}
The measurement associated with the subsequent readout of the meter is weak in the sense that it leaves the system undisturbed to first order in $\epsilon$. This type of measurement can thus be performed unnoticed, and our formalism should enable prediction of its outcome in a seemingly "counter-factual" manner ("If $\hat{A}$ had been measured, the outcome would have been ... ").
The nature of the weak measurement allows an unknown agent to perform measurements without our knowledge and hide the result from us.
At any later time, this person can inform us of the result, and we need the present theory to most accurately predict the outcome of this result.
This ability comes with a cost, as to the interpretation of the weak measurement result, which we will return to shortly.

An arbitrary projective measurement of the meter is described by linear combinations of $\hMeter_\downarrow$ and $\hMeter_\uparrow$. For measurements in the $\hat \sigma_x$ and $\hat \sigma_y$-bases we can express the conditional expectation of $\hat \sigma_x = (\hat \sigma + \hat \sigma^\dagger)$ and $\hat \sigma_y = i(\hat \sigma - \hat \sigma^\dagger)$ by the real and imaginary parts of the expectation value of the step down operator $\hat\sigma$, respectively.

The expectation value of the meter operator $\hat\sigma$ is
\begin{align}
 \Expectation[\hat\sigma | y] =
 \frac{ \sum_{k=1}^K \trace(\hSystem_{k|y} \epsilon \hat{A} \rho_0 \hSystem_{k|y}^\dagger)  }
 { \sum_{k=1}^K \trace\left( \hSystem_{k|y} \rho_0 \hSystem_{k|y}^\dagger \right) }.
\end{align}
In this formula the denominator is independent of $\hat A$. This is due to the weak nature of the measurement, and it
implies, that the expectation value is \emph{linear} in the system operator $\hat A$. Since the meter couples to the system observable $\hat{A}$, it is natural to consider the weak value,
\begin{align}
 \expectedweak{A} \equiv
 \lim_{\epsilon\to 0} \frac{1}{\epsilon} \Expectation[\hat \sigma|y] = \frac{ \sum_{k=1}^K \trace(\hSystem_{k|y} \hat A \rho_0 \hSystem_{k|y}^\dagger)  }
 { \sum_{k=1}^K \trace\left( \hSystem_{k|y} \rho_0 \hSystem_{k|y}^\dagger \right) }. \label{eq:PastDensityMatrixDefinition}
\end{align}

We observe that to retrieve the weak value, i.e., the average outcome of a weak measurement, the information in the past quantum state $\Xi=(\rho_0,E)$ with the  effect matrix $E = \sum_{k=1}^K \hSystem_{k|y}^\dagger \hSystem_{k|y}$ can be deduced from a \emph{past density matrix} $\rho_\past$.
The expression $\expectedweak{A} =  \trace(\hat A\rho_\past)$ holds if we identify
\begin{align}
 \rho_\past = \frac{\rho_0 E}{\trace(\rho_0 E)}. \label{eq:PastDensityMatrixProduct}
\end{align}

It is worth noting that $\rho_\past$ is not Hermitian and even if $\hat{A}$ is Hermitian, $\trace(\hat A\rho_\past)$ may have both real and imaginary parts.
This is a well-known property of weak measurements, and it does not require the readout of, non-physical, complex measurement results, since the real part of $\expectedweak{A}$ refers to the (real) mean value of the $\hat \sigma_x$-operator of the meter, while its imaginary part is obtained by measuring the average $\hat \sigma_y$-spin component of the meter.
These averages may, in turn, indicate mean values of the system observable that are very different from its spectrum of eigenvalues.
But this is well understood as an interference effect, when a measurement outcome is conditioned on different pre- and post-selected state of a physical system. \cite{Aharonov.PhysRevLett.60.1351(1988)}

\section{Projective read-out measurements and the past quantum state}

Imagine now that we perform a projective read-out measurement of the system observable $\hat A$ using our meter.
In this case $\hMeter_m = \hat \Pi_{a_m}$ where the different measurement results are the eigenvalues of the observable $\hat A$ denoted $a_m$.
Following \refeq{eq:ConditionalMeterProbabilityState} the probability for observing the eigenvalue $a$ conditional on the later measurement result $y$ is
\begin{align*}
 P(a|y) = \frac{\trace(\hat\Pi_{a_m} \rho_{0-} \hat\Pi_{a_m} E_+)}{\trace\left(\sum_{m'} \hat\Pi_{a_{m'}} \rho_{0-} \hat \Pi_{a_{m'}} E_+\right)}
\end{align*}
where we by the $+$ and $-$ signs emphasize, that $E_+$ is the effect matrix including measurements from immediately \emph{after} the projective read-out measurement was performed until time $T$ (here exemplified with a single measurement with result $y$), and $\rho_{0-}$ is the usual quantum state conditioned on the measurements until immediately \emph{before} the projective read-out measurement was performed.
The expectation value of the projective measurement of the observable $\hat A$ is then
\begin{align*}
 \Expectation[\hat A|y] = \sum_m a_m P(a_m|y) = \frac{\trace(\sum_m a_m \hat\Pi_{a_m} \rho_{0-} \hat\Pi_{a_m} E_+)}{\trace\left(\sum_{m'} \hat\Pi_{a_{m'}} \rho_{0-} \hat \Pi_{a_{m'}} E_+ \right)}.
\end{align*}
By inserting a resolution of the identity $\id = \sum_{m'} \hat\Pi_{a_{m'}}$ this expression can be written in two ways
\begin{align*}
 \Expectation[\hat A|y]
  = \frac{\trace\left( \hat A \rho_{0-} \sum_m \hat \Pi_{a_m} E_+ \hat \Pi_{a_m}\right)}{\trace\left(\rho_{0-} \sum_{m} \hat \Pi_{a_m} E_+ \hat\Pi_{a_m} \right)}
  = \frac{\trace\left( \hat A \sum_m \hat \Pi_{a_m} \rho_{0-} \hat \Pi_{a_m} E_+ \right)}{\trace\left(\sum_{m} \hat \Pi_{a_m} \rho_{0-}  \hat\Pi_{a_m} E_+  \right)}.
\end{align*}
By introducing the effect matrix which takes the unobserved projective measurement into account by the map, $E_- = \sum_m \hat \Pi_{a_m} E_+ \hat \Pi_{a_m}$ we can write this result
\begin{align}
 \Expectation[\hat A|y] = \trace(\hat A \rho_{\mathrm p-}),
 \label{eq:ProjectivePriorMeasurement}
\end{align}
defining the past density matrix as $\rho_{\mathrm p-} = \rho_{0-}E_-/\trace(\rho_{0-}E_-)$ immediately prior to the projective measurement.

Alternatively, we get
\begin{align}
 \Expectation[\hat A|y] = \frac{\trace(\hat A \rho_{0+} E_+)}{\trace(\rho_{0+} E_+)}, \label{eq:ProjectiveAfterMeasurement}
\end{align}
by using the past density matrix $\rho_{\mathrm p+} = \rho_{0+}E_+/\trace(\rho_{0+}E_+)$, with the unobserved measurement process modifying the state $\rho_{0+} = \sum_m \hat\Pi_{a_m} \rho_{0-} \hat\Pi_{a_m}$.

The projective measurement disturbs the system, and it must hence be taken into account in one of the components of the past quantum state: In \refeq{eq:ProjectivePriorMeasurement} it is included in the effect matrix $E_-$ whereas in \refeq{eq:ProjectiveAfterMeasurement} it is included in the forward density matrix $\rho_{0+}$. The two resulting alternative forms for the expectation value of $\hat{A}$ are equivalent since $\hat A$ commutes with the effect of the projective measurement.

While the general formula \refeq{eq:ConditionalMeterExpectation} using the past quantum state $\Xi = (\rho_0, E)$ can be applied to yield the probabilities for any past measurement outcome, it is interesting that the formalism may be brought closer to usual mean value expressions by use of the appropriately defined past density matrix.

\section{Differential equations for homodyne and counting measurements} \label{sec:QuantumSmoothingEquations}

A quantum system subject to homodyne or heterodyne detection satisfies an Itô stochastic differential equation of the form given in Eq.~(3).
As discussed previously such a measurement scenario fits into the present formulation by using the measurement operator for the time interval from $t$ to $t + dt$ is given by \refeq{eq:DiffusionMeasurementOperator} if we assume 100\% efficiency and no unobserved channels, i.e. no dissipation effects.
The adjoint update \refeq{eq:AdjointOperation} is then
\begin{align}
 dE_t \equiv E_{t - dt} - E_t = \left[ i\cm{\hat H, E_t } - \frac{1}{2}\acm{\hat c^\dagger \hat c, E_t } + \hat c^\dagger E_t \hat c \right] dt
 + \left[ \hat c^\dagger E_t + E_t \hat c \right] dY_{t - dt}. \label{eq:HomodyneBackwards}
\end{align}
where we have chosen a normalization of the effect matrix $E$ such that the front factor $(2\pi dt)^{-1/4} \exp(-dY^2/4 dt)$ is discarded and $E(T) = \id$ where $T$ is the final time of observation.

By including dissipation effects and limited detector efficiency we obtain Eq.~(4).

The conditional time evolution of a quantum system state subject to discrete counting signals in $N$ channels can be described by the infinitesimal operators
\begin{align}
\begin{aligned}
 \hSystem_0 &= \id - i \hat H dt - \sum_{m=1}^N \frac{1}{2} \hat L_m^\dagger \hat L_m dt \\
 \hSystem_m &= \hat L_m \sqrt{dt} \qquad \text{for $1 \leq m \leq N$}
\end{aligned}
\end{align}
where $\hat L_m$ are Lindblad operators describing quantum jumps associated with the emission of quanta by the system into the environment.
$\hSystem_0$ yields the measurement effect when no quantum (photon) is detected and $\hSystem_m$ indicates that a quantum is detected in environment channel number $m$.

If only the first channel is detected we get the well-established quantum jump filtering equation\cite{Wiseman.QuantumMeasurementControl}
\begin{multline}
 d\rho_t = \left[ -i\cm{\hat H, \rho_t} + \sum_{m=2}^N \left( \hat L_m \rho_t \hat L_m^\dagger - \frac{1}{2}\acm{\hat L_m^\dagger \hat L_m, \rho_t} \right) - \frac{1}{2}\acm{\hat L_1^\dagger \hat L_1, \rho_t} + \trace(\hat L_1^\dagger \hat L_1 \rho_t) \rho_t \right] dt \\
 + \left[ \frac{\hat L_1 \rho_t \hat L_1^\dagger}{\trace(\hat L_1^\dagger \hat L_1 \rho_t)} - \rho_t \right] dN_t,
\end{multline}
where $dN_t = 0$ in all the intervals where no photon is detected, but $dN_t = 1$ (and $dt = 0$) at the instants of time a photon is detected in channel 1.

The adjoint update of the effect matrix is
\begin{multline}
 dE_t \equiv E_{t-dt} - E_t = \left[ i\cm{\hat H, E_t } + \sum_{m=2}^N \left( \hat L_m^\dagger E_t \hat L_m - \frac{1}{2}\acm{\hat L_m^\dagger \hat L_m, E_t} \right) - \frac{1}{2}\acm{\hat L_1^\dagger \hat L_1, E_t} \right] dt \\
 + \left[ \hat L_1^\dagger E_t \hat L_1 - E_t \right] dN_t.
\end{multline}

\section{Time evolution of mean values and weak values}
For a quantum system subject to unit efficiency homodyne detection and no unobserved dissipation channels, the conventional mean value of a system observable
$\langle \hat{A}\rangle = \trace(\hat{A}\rho_t)$ changes with time according to the stochastic differential,
\begin{multline} \label{noisemean}
 d\braket{\hat A} = -i\trace(\cm{\hat A, \hat H}\rho_t) dt + \trace\left( \hat A \hat c \rho_t \hat c^\dagger - \frac{1}{2}\acm{\hat A, \hat c^\dagger \hat c} \rho_t \right) dt \\
  + \left( \trace( (\hat A \hat c + \hat c^\dagger \hat A)\rho_t ) - \trace((\hat c+ \hat c^\dagger)\rho_t) \trace(\hat A\rho_t) \right) dW_t,
\end{multline}
where $dW_t = dY_t - \trace((\hat c+ \hat c^\dagger)\rho_t) dt$. 
 
We now address the similar change according to the past quantum state, i.e, the change of the weak value estimate $\langle \hat{A}\rangle_w=\trace(\hat A\rho_{\mathrm p})$. The differential of the un-normalized past density matrix is
\begin{align*}
 d(\rho_t E_t) &= \rho_{t+dt} E_{t+dt} - \rho_t E_t = d\rho_t E_{t+dt} - \rho_t dE_{t+dt}.
\end{align*}
By inserting the expressions \refeq{eq:HomodyneBackwards} and Eq.~(3) we get
\begin{multline}
 d(\rho_t E_t) = -i\cm{\hat H, \rho_t E_{t+dt}} dt + \cm{\hat c, \rho_t \hat c^\dagger E_{t+dt}} dt - \frac{1}{2}\cm{\hat c^\dagger \hat c, \rho_t E_{t+dt}} dt
  + \cm{\hat c, \rho_t E_{t+dt}} dY_t.
\end{multline}
Note that $\trace(d(\rho_t E_t)) = 0$ as expected since $\trace(\rho_t E_t) = \trace(\rho_T)$ is constant.
The differential of the weak value $\expectedweak{\hat A}$, $d\expectedweak{\hat A} = d\trace(\hat A\rho_{\mathrm{p},t}) = \trace(\hat A d(\rho_t E_t)) / \trace(\rho_t E_t)$ thus becomes
\begin{multline}
 d\expectedweak{\hat A} = \frac{1}{\trace(\rho_t E_t)} \left[ -i\trace([\hat A,\hat H]\rho_t E_{t+dt})dt \right.
 + \trace\left( \cm{\hat A,\hat c}\rho_t \hat c^\dagger E_{t+dt} - \frac{1}{2} \cm{\hat A, \hat c^\dagger \hat c} \rho_t E_{t+dt}\right) dt \\
 + \left. \trace([\hat A,\hat c] \rho_t E_{t+dt})dY_t \right].
\end{multline}

This differs from the change in the conventional mean value and, in particular, we observe the suppression of the noise term $\propto dY_t$ for observables which commute with the
measurement operator $\hat c$. This suppression explains the smooth behavior of the black dashed curves in Figs.~2(a) and 2(b), while the conventional mean values $\langle \hat{A}\rangle$ show fluctuations due to the  Wiener noise term, $dW$, in Eq.~(\ref{noisemean}).

\chapter{Relation to classical hidden Markov models}
The classical theory of hidden Markov models is very well established and an excellent introduction can be found in Ref.~\onlinecite{NumericalRecipesThirdEd}.
Here, a \emph{hidden Markov model} is a discrete-time stochastic process where a hidden system state evolves according to a Markov chain and observations of some output signal depending on the system state are performed.
Let $X_t$ be the system state at time $t$ and the output signal at time $t$ be $Y_t$.
The output signal $Y_t$ depends only on the system state at the same time and its probability distribution is therefore completely determined by $P(Y_t|X_t)$.
Since the evolution of $X_t$ follows a Markov chain the system dynamics is completely determined by the transition probabilities $P(X_{t+1}|X_t)$.
It is a simple matter to generalize these ideas to the continuous time case (such as a system governed by a rate equation), but for simplicity we will consider only the discrete time case here and the times $t$ are integers.

The full joint probability distribution for measurements $Y_t$ and states for a process running from time $t = 0$ to time $t = T$ is then
\begin{align}
 P(X_0, \ldots, X_T, Y_1, \ldots Y_T) = \prod_{t=0}^{T-1} P(X_{t+1}|X_t) \prod_{t=1}^T P(Y_t|X_t) P(X_0), \label{eq:HMMJoint}
\end{align}
where $P(X_0)$ is the probability for the initial state $X_0$.
The theory of hidden Markov models provides numerically efficient algorithms for calculating the following quantities: (1) the forward filtered state probability distribution $P(X_t|Y_1, \ldots Y_t)$ and (2) the smoothed state probability distribution $P(X_t|Y_1, \ldots Y_T)$ for all times $t$.

The forward estimate is easily calculated by a standard recursive Bayesian procedure.
Following the notation in \cite{NumericalRecipesThirdEd} we define the vectors
\begin{align}
 \alpha_t(i) &= P(Y_1, \ldots Y_t, X_t = i)  \label{eq:HMMAlphaDefinition} \\
 \beta_t(i) &= P(Y_{t+1}, \ldots Y_N | X_t = i). \label{eq:HMMBetaDefinition}
\end{align}
Using the $\alpha_t$ and $\beta_t$-vectors we can calculate the filtered and smoothed distributions at time $t$ by the following formulas
\begin{align}
 P(X_t = i| Y_1, \ldots Y_t) &= \frac{\alpha_t(i)}{\sum_k \alpha_t(k)}, \label{eq:HMMFilteredEstimate} \\
 P(X_t = i| Y_1, \ldots Y_N) &= \frac{\alpha_t(i) \beta_t(i)}{\sum_k \alpha_t(k) \beta_t(k)}, \label{eq:HMMSmoothedEstimate}
\end{align}
both of which are variations of Bayes' formula.
It is not difficult to show that $\alpha_t$ and $\beta_t$ satisfy the following recursion relations
\begin{align}
  \alpha_{t+1}(i) &= \sum_{j} P(Y_{t+1}|X_{t+1} = i ) P(X_{t+1} = i |X_t = j) \alpha_t(j) \label{eq:AlphaUpdate} \\
  \beta_t(i) &= \sum_j P(Y_{t+1}| X_{t+1} = j) P( X_{t+1} = j | X_t = i ) \beta_{t+1}(j), \label{eq:BetaUpdate}
\end{align}
where $\beta_T(i) = 1$ and $\alpha_0(i) = P(X_0 = i)$.
In the following, we show that our effect matrix is equivalent to the $\beta_t$-vector and that the past quantum state $\Xi(t)$ is the natural quantum generalization of the hidden Markov model pair $(\alpha_t, \beta_t)$.
In the case of non-disturbing measurements the hidden Markov model smoothed state (\ref{eq:HMMSmoothedEstimate}) is equivalent to both the past quantum state $\Xi$ and the past density matrix $\rho_\mathrm{p}$. This is consistent with the weak value assumption of no disturbance of the system due to the measurements since classical measurements may always be thought of as non-disturbing).

Indeed, the above hidden Markov theory can be formulated using diagonal density matrices and updates of the form given in \refeq{eq:QuantumOperation}.
Let $\ket i$ be an orthogonal basis for a Hilbert space, where $i$ denotes the same internal states as in the hidden Markov model.
The Markov chain evolution $P(X_{t+1}|X_t)$ is now given by the update
\begin{align}
 \mathcal C\colon \rho \longmapsto \sum_{i,j} \ket{j}\bra{j}  P(X_{t+1} = j|X_t = i) \braket{i|\rho|i},
\end{align}
which is in fact an evolution of the type given in \refeq{eq:QuantumOperation} with $\hSystem_{i,j} = \sqrt{P(X_{t+1} = j |X_t = i)} \ket j \bra i$
The observation process is given by the update
\begin{align}
 \mathcal I \colon \rho \overset{y}{\longmapsto} \sum_i P(Y_t = y|X_t = i)\braket{i|\rho|i} \ket i \bra i
\end{align}
followed by re-normalization as in \refeq{eq:QuantumOperation}.

The classical hidden Markov model is then reproduced by picking the initial state $\rho_0 = \sum_i P(X_0 = i) \ket i \bra i$ and applying the two updates $\mathcal C$ and $\mathcal I$ in sequence.

The (un-normalized) forward filtered state $\alpha_t$ is simply the (un-normalized) filtered quantum state $\tilde\rho_t$ which satisfies the recursion relation
\begin{align}
 \tilde\rho_t &\overset{\mathcal C}{\longmapsto} \sum_{i,j} \ket{i}\bra{i}  P(X_{t+1} = i|X_t = j) \braket{j|\tilde\rho_t|j}  \\
 & \overset{\mathcal I}{\longmapsto} \sum_{i,j} \ket{i}\bra{i} P(Y_{t+1} = y|X_{t+1}=i)  P(X_{t+1} = i|X_t = j) \braket{j|\tilde\rho_t|j} \equiv \tilde\rho_{t+1}
\end{align}
which for diagonal $\tilde\rho_t$ is exactly \refeq{eq:AlphaUpdate}.

The effect matrix $E_t$ is initially $E_T = \id$ which is equivalent to the initial condition $\beta_T(i) = 1$.
The effect matrix is propagated according to the adjoint update \refeq{eq:AdjointOperation} as
\begin{align}
 E_{t+1} &\overset{\mathcal I^\dagger}{\longmapsto} \sum_j \ket j \bra j P(Y_{t+1} = y|X_{t+1} = j) \braket{j|E_{t+1}|j}  \\
 &\overset{\mathcal C^\dagger}{\longmapsto} \sum_{i,j}\ket i\bra i P(Y_{t+1} = y|X_{t+1} = j) P(X_{t+1} = j|X_t = i)  \braket{j|E_{t+1}|j} \equiv E_t.
\end{align}
which is exactly the update formula for $\beta_t$ \refeq{eq:BetaUpdate}.
Note here that the $\mathcal I$-update
is unchanged, whereas the adjoint update for $\mathcal C$ is modified.

The theory of hidden Markov models includes numerically efficient algorithms for re-estimating the parameters occurring in the model.
The so-called Baum-Welch algorithm is a special case of the Expectation-Maximization algorithm.
By using the fully smoothed estimate as encoded in the $\alpha$- and $\beta$-vectors, a simple formula for the re-estimated parameters exists, which leads to a more likely sequence of measurement results $Y_1, \ldots Y_T$ given the model.
In this way a local maximum of the likelihood can be calculated by iterating parameter re-estimation and the smoothing calculation.
We believe that a similar technique can be applied to estimate unknown parameters in quantum processes using the past quantum state.


%